\documentclass[prl,aps,showkeys,twocolumn]{revtex4}
\usepackage{epsfig}
\usepackage{graphicx}
\usepackage{amssymb}
\usepackage{color}

\begin{document}

\title{Symmetry assisted preparation of entangled many-body states on a quantum computer}

\author{Denis Lacroix } \email{denis.lacroix@ijclab.in2p3.fr}
\affiliation{Universit\'e Paris-Saclay, CNRS/IN2P3, IJCLab, 91405 Orsay, France}

\date{\today}
\begin{abstract}
Starting from the Quantum-Phase-Estimate (QPE) algorithm, a method is proposed to construct entangled states 
that describe correlated many-body systems on quantum computers. Using operators for which the discrete set of eigenvalues is known, the QPE approach is followed by measurements that serve as projectors on the entangled states. These states can then be used as inputs for further quantum or hybrid quantum-classical processing. When the operator is associated to a symmetry of the Hamiltonian, the approach can be seen as a quantum--computer formulation of symmetry breaking followed by symmetry restoration. The method,  called Discrete Spectra Assisted (DSA), is applied to superfluid systems. By using the blocking technique adapted to qubits, the full spectra of a pairing Hamiltonian is obtained.         
\end{abstract}


\maketitle

The development of quantum devices with increasing numbers of qubits is nowadays experiencing  rapid
and exciting progress. This opens new perspectives to solve complex problems that are out of reach of classical 
computers \cite{Nie02,Hid19}. The simulation of complex quantum systems, such as many-body 
interacting fermions, appears as one of the perfect playground where quantum computing 
can lead to a significant boost. Quite naturally,  an increasing number of innovative methods are now proposed  
to describe this problem on an ensemble of qubits.  
In recent years, the number of applications, sometimes on real quantum 
devices, is increasing rapidly not only in quantum chemistry \cite{Mcc17,Fan19,Cao19,McA20,Bau20,Lan10,Bab15,OMa16,Col18,Hem18} but also in condensed matter \cite{Mac18}, nuclear physics \cite{Dum18,Lu19,Rog19,Du20}, and in quantum field theories \cite{Klc18,Klc19,Ale19,Lam19}. 

The use of quantum computers requires often to reinvent techniques that are standardly 
used in classical devices. Among the standard techniques widely used in mesoscopic systems, the possibility 
to use symmetry breaking (SB) trial wave-functions followed by proper symmetry restorations (SR) allows for including correlations beyond the perturbative regime. The SB-SR strategy is for example a pillar in the treatment of the nuclear many-body
problem where the number of constituents varies from very few to several hundreds \cite{Rin80,Bla86,Ben03,Rob18}. 
While the first step (SB) can be seen as a simplification to grasp correlations, the second step (SR)  is much more demanding. In nuclear physics, 
the use of trial wave-packets after projection in a variational principle (Variation After Projection) is at the forefront of current 
capabilities of classical computers, especially if several symmetries like particle number, angular momentum, ... are simultaneously broken \cite{Ben03, Rob18}.   
The many-body state-vectors after projection correspond to highly entangled state. Entangled states are building blocks of many algorithms used in quantum computing \cite{Nie02,Hid19} and it is quite natural to investigate if these states can be accurately obtained/manipulated with a quantum computer. I propose here a methodology to prepare strongly entangled states based on the SB-SR strategy using the Quantum-Phase-Estimate (QPE) method. 

\begin{figure}[htbp]  
\includegraphics[width= 1.0\linewidth]{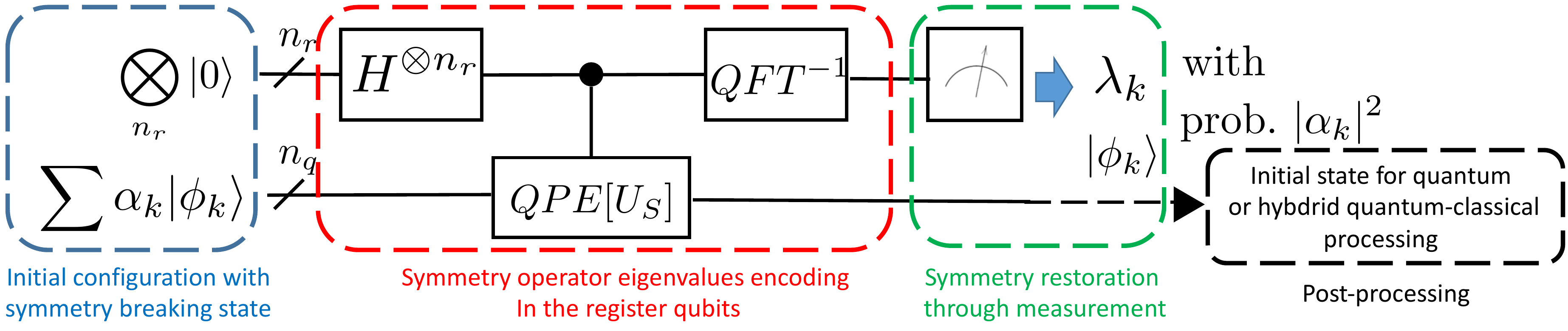}
    \caption{ Illustration of the protocol proposed here to prepare a strongly entangled state in a quantum computer using the DSA method. The starting configuration (blue box)  is a state $\sum_k \alpha_k |\phi_k \rangle$ and a set of register qubits. The QPE is applied using an operator $U_S$ with known discrete spectra (red box). The measurement of register qubits acts as a projector on a set of 
    entangled states that are eigenvectors of the operator $U_S$ (green box). The state can then be used for post-processing (black box). }
    \label{fig:protocol}
\end{figure}
        
The QPE approach that is based on the Quantum Fourier Transform (QFT) \cite{Nie02,Hid19} is a practical way to obtain on a quantum computer 
estimates of the eigenvalues of a unitary operator $U$ acting on $n_q$ qubits. This approach makes use of a set of $n_r$ register qubits that couple to the working qubits 
through a repeated applications of controlled-$U$ gates. Denoting by $e^{i{2\pi} \theta_k}$ a given eigenvalue of $U$, the QPE approach returns an approximation of the phase, $\tilde \theta_k$, written as a truncated binary fraction whose precision to describe $\theta_k$ depends on $n_r$. The QPE is well documented \cite{Nie02}, and I only give in Fig. \ref{fig:protocol} a schematic view of the QPE 
quantum circuits (additional discussions on QPE can be found in \cite{Fan19, Ovr03,Ovr07}). I assume that the initial state $| \psi \rangle$ is written in the $n_q$ qubits and decomposes as $| \psi \rangle = \sum_k  \alpha_k | \phi_k \rangle $ where $| \phi_k \rangle$ are eigenvectors associated to the set of phases $\theta_k$. After the inverse QFT,  the state denoted by $| \psi_f \rangle$ becomes: 
\begin{eqnarray}
| \psi_f \rangle &=& \sum_k \alpha_k | \theta_k 2^{n_r} \rangle \otimes  | \phi_k \rangle .  \label{eq:phifinal}
\end{eqnarray}  
Here, $| \theta_k 2^{n_r} \rangle$ should be understood as a binary string of $0$ and $1$ that corresponds to the binary fraction of $\theta_k$ truncated at the $1/2^{n_r}$ term.  The eigenvalue estimates 
are obtained through repeated measurements of the registered qubits. In first approximation\footnote{In practice, since the eigenvalues can rarely be written exactly  as a truncated binary fraction and since quantum computers 
are not ideal, a set of surrounding binary strings are also measured \cite{Nie02, Ovr07}.},
 the binary number $\{ \theta_k 2^{n_r} \} $   is obtained with a probability 
 $|\alpha_k|^2$.  After the measurement, the state is projected on one of the channels 
$| \theta_k 2^{n_r} \rangle \otimes  | \phi_k \rangle$.  

In the present work, I propose to use the QPE approach for operators with already known eigenvalues in order to obtain 
strongly entangled states that are difficult to construct on a classical computer.  
A hermitian operator $S$ acting on the $n_q$ qubits is considered. This operator has a finite discrete set of eigenvalues  written in ascending order as $\{\lambda_0 \le  \cdots \le  \lambda_M\}$.  
I illustrate first the approach to a specific situation where the set of eigenvalues can be connected to a set of  integers $\{m_0 \le  \cdots \le  m_M\}$ through a linear 
relation $\lambda_k = a m_k$, where $a$ is a constant. This example is particularly important because it includes operators that  are linked to symmetries such as parity, particle number or angular momentum  operators. The generalization to a more general class of operators is discussed below. 
For the restricted class of operator, I define the unitary operator $U_S$ as:
\begin{eqnarray}
U_S &=& \displaystyle  \exp\left\{ 2\pi i \left[ \frac{S - \lambda_0}{  a 2^{n_0} } \right] \right\} .  \label{eq:us}
\end{eqnarray} 
The phase associated to each eigenvalue of $U_S$ is given by $\theta_k = (m_k - m_0)/2^{n_0}$.  Imposing $\theta_k<1$ for all eigenvalues 
leads to the condition $n_0> \ln(m_k - m_0)/\ln 2$. $\theta_k$  is then automatically  exactly written as a binary fraction truncated at the $n_0$ 
term.  When applying the QPE approach for $S$, an optimal choice for the number of register qubits is $n_r=n_0$ with:
\begin{eqnarray}
n_r - 1 \le \ln(m_k - m_0)/\ln 2 < n_r.
\end{eqnarray}   
In the following applications, the lowest (optimal) value of $n_r$ for which the conditions are verified  is used as well as $n_r=n_0$. Taking higher values of $n_r$ will lead to useless register qubits. Lower values are a priori possible but this will degrade the selectivity of the states after the measurement. With this optimal choice, the binary strings entering in the registered components of 
Eq. (\ref{eq:phifinal}) are directly those corresponding to the $(m_k - m_0)$ values. 

The specific choice of $U_S$ given by Eq. (\ref{eq:phifinal}) with the optimal value of $n_r$ is particularly suitable for selecting the component $| \phi_k \rangle$ associated to the eigenvalue $m_k$\footnote{Note that eigenvalues can also be degenerated. In this case, the state should be understood as a quantum mixing of different eigenstates for the degenerated eigenvalues.}. Indeed, since the phases $\theta_k= (m_k - m_0)/2^{n_r}$  exactly write as truncated binary fractions, there is no pollution from other contributions in an ideal quantum device. The ultimate goal of the approach
is to obtain  after measurements the set of states $| \phi_k \rangle$. These states might have highly nontrivial properties depending on the choice of the operator $S$.  They can then be used in a second step for further quantum and/or hybrid processing like in the Variational Quantum Eigenvalue (VQE) method \cite{Per14,McC16}. Since the present approach is based on the use of known discretized spectra 
for specific operators, I call it Discrete Spectra Assisted (DSA) approach in the following. 
I illustrate below that symmetry restoration can 
be achieved using this technology. Interesting discussions on symmetry restoration within quantum computers can be found in Ref. 
\cite{Whi13,Tsu20,Rya18,Yen19,Mol16,Gar19}. 
The full protocol proposed here is illustrated in Fig. \ref{fig:protocol}.  
An initial state 
with some broken symmetry is prepared on the working qubits. An operator $S$ associated to the symmetry we aim to restore is then chosen 
and the QPE is applied to $U_S$. The register qubits repeated measurements lead to a set of states that respect the symmetry. The last step replaces the 
symmetry restoration process. 
   
The methodology 
is illustrated here for the $U(1)$ symmetry associated to particle number.  
This symmetry breaking, used in the BCS or Hartree-Fock Bogolyubov (HFB) theories, 
is particularly powerful to account for superfluidity but a precise description of finite systems 
can only be achieved once the symmetry is restored \cite{Bla86,Von01,Zel03,Duk04,Bri05}. The goal here is to describe 
many-body systems, it is then convenient to introduce a single-particle basis associated to creation/annihilation operators 
$(a^\dagger_j, a_j)$. The mapping between single-particle states to qubits is made by using the Jordan-Wigner Transformation (JWT) \cite{Jor28,Lie61,Som02,See12, Dum18, Fan19}
with the convention: 
\begin{eqnarray}
a^\dagger_j &\longrightarrow &  Q^+_j \otimes Z^{<}_{j-1} , \label{eq:jwt}
\end{eqnarray}
where $Q^+_j = \frac{1}{2} \left( X_j - i Y_j \right) $ and $Z^{<}_{j-1} =  \bigotimes_{k=1}^{j-1} (-Z_k)$.  $(X_j$, $Y_j$, $Z_j)$  together with the 
identity $I_j$ 
are the standard unary gates applied to the qubit $j$.  A natural choice for $S$ for the $U(1)$ symmetry is to take the equivalent to the particle number operator $\hat N = \sum_i a^\dagger_i a_i$. This operator, counts the number of occupied qubits in the $n_q$ basis. With the convention (\ref{eq:jwt}), it is given by $\hat N = \sum_j Q^+_j Q_j = \frac{1}{2} \sum_{j}(I_j-Z_j)$. The operator defined in Eq. (\ref{eq:us}) is denoted simply as $U_N$ below. It can be decomposed as a product of operators acting on each qubit $U_N = \prod_j U_j$
where $U_j$ acts on the qubit $j$ \footnote{The operator $U_N$ is diagonal in the working qubit basis. For a given element of this basis, we simply have: 
$U_N  | \delta_{n_q-1}, \cdots , \delta_{0} \rangle = e^{i\pi [\sum_j \delta_j ]/2^{n_0-1}} | \delta_{n_q-1}, \cdots , \delta_{0} \rangle$
where all $\delta_j$ are $0$ or $1$.}  and is given by $U_i = | 0_i \rangle \langle  0_i | + \exp(i\pi/2^{n_0-1}) | 1_i \rangle \langle  1_i |$.  

The application of the methodology with $U_N$ gives access to the probability distribution of the number of occupied qubits in the initial 
state $| \psi \rangle$, the so-called counting statistics  in many-body systems (see for instance \cite{Lac20} and ref. therein). For qubits, I call this distribution qubit counting statistics (QCS). Note that the components 
of the registered qubits prior to he inverse QFT (see Fig. \ref{fig:protocol}) give access to the generating function of the QCS \cite{Lac20}. 

\begin{figure}[htbp]  
\includegraphics[width= 0.95\linewidth]{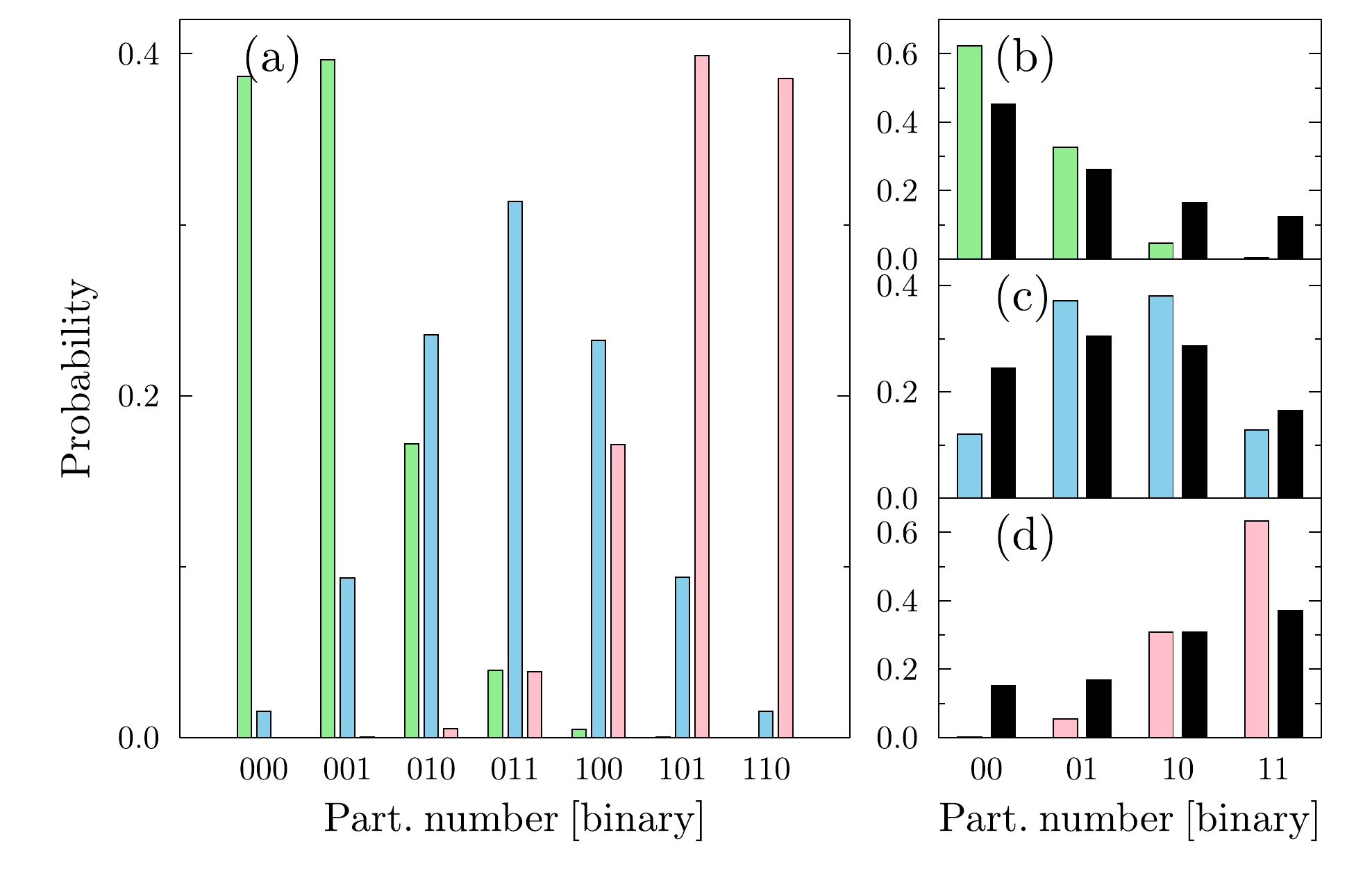} 
    \caption{(a) Illustration of the QCS obtained using the DSA approach with the $U_N$ operator for the state $| \psi \rangle = \bigotimes_{n_q} R^j_Y(\varphi) | 0_j \rangle$ for $n_q=6$ and with $\varphi=\pi/4$ (green),  $\varphi=\pi/2$ (blue) and $\varphi=3\pi/4$ (pink). 
    The $x$-axis corresponds to the binary fraction $A/2^{n_r}$ where $A$ is the particle number. In this illustration, 
    $n_r=3$ and for instance $A=6$ particles give  $6/8 = 1/2 +1/4 +0/8 \equiv [110]$. 
In panels (b), (c) and (d) the results of the DSA method obtained with the Qiskit software are shown for $n_q=3$ and $n_r=2$  for $\varphi=\pi/4$,  $\varphi=\pi/2$  and $\varphi=3\pi/4$, respectively. In these panels, the results obtained using the IBM Q 5-qubits  {\sl 'ibmq\_vigo'} real device with 2048 events are systematically shown in black. }
    \label{fig:proj1precise}
\end{figure}

As a first illustration, some QCS obtained numerically with the IBM Qiskit  toolkit \cite{Abr19} using the protocol of Fig. \ref{fig:protocol} are shown in Fig. \ref{fig:proj1precise}. In these examples, the initial states are obtained from a coherent $Y$-rotation of all working qubits with
 $| \psi \rangle = \bigotimes_{n_q} R^j_Y(\varphi) | 0_j \rangle$, where $R^j_Y(\varphi)=e^{-i\varphi Y_j/2}$. In these ideal calculation, I numerically obtained 
 that the 
 QCS probability $P(A)$ to have $A$ occupied qubits in the $n_q$ qubits properly identifies with $P(A) = C^A_{n_q} p^{A} (1-p)^{n_q -A}$ where $p=\sin^2(\varphi/2)$. 
 
In panels (b)-(c) of Fig. \ref{fig:proj1precise}, results of the DSA method obtained with a real 5-qubit quantum device are also shown in blackand compared to the ideal case. Deviations from the ideal quantum emulator case sign the effect of noise. 
Nevertheless, the fact that the trends in the QCS are globally reproduced is rather encouraging for the future use of the approach in the NISQ (Noisy Intermediate-Scale Quantum) context. Note that, I also performed tests on the 15-qubit device provided in IBM Q and the results were essentially compatible with a white noise. 

With the aim of (i) validating the full method including the post-processing after measurement 
and (ii) illustrating the powerfulness of the approach, I apply the technique to describe 
a set of $n_q$ fermions interacting through the pairing Hamiltonian \cite{Von01,Zel03,Duk04,Bri05}:      
\begin{eqnarray}
H_{\rm P} &=& \sum_{i>0} \varepsilon_i (a^\dagger_i  a_i + a^\dagger_{\bar i} a_{\bar i}) - g \sum_{i,j>0} 
a^\dagger_i a^\dagger_{\bar i}  a_{\bar j} a_j . \label{eq:hpair}
\end{eqnarray} 
$(i,{\bar i})$ denotes a pair of time-reversed states, and $i>0$ means that summations are made on pair labels.  Note 
that here $(i,\bar i)$ labels pair of states in the Fermion Fock space. The Hamiltonian is highly non-local since each pair interacts with all other pairs. It 
was already considered in Ref. \cite{Ovr07} for quantum computation using the standard QPE technique.  
 The Hilbert space is mapped to a set of qubits $n=1, \cdots, n_q$
using the JWT technique. By convention, it is assumed here that if $i$ is described by the qubit $n$, then its time-reversed state ${\bar i}$ is described by the qubit $n+1$, such that $a^\dagger_i a^\dagger_{\bar i} \rightarrow Q^+_n Q^+_{n+1}$.  

I  consider below the degenerate case $({\varepsilon_i} = {\varepsilon} = 0)$ for which the energy of the eigenstates with $A$ particles is known analytically and is given 
by \cite{Bri05}:
\begin{eqnarray}
E/g = -\frac{1}{4} (A-\nu) (2 n_q - A - \nu - 2). \label{eq:senior}
\end{eqnarray}     
This equation holds for odd or even particle numbers. 
$\nu$ denotes the number of broken pairs $(i, \bar i)$  in the eigenstates, that is the so-called seniority (for more details on the seniority see for instance \cite{Rin81,Bri05}).
Increasing the values of $\nu$ gives access to the different excited-state energies 
for a fixed $A$. 
\begin{figure}[h]  
\includegraphics[width= 0.7\linewidth]{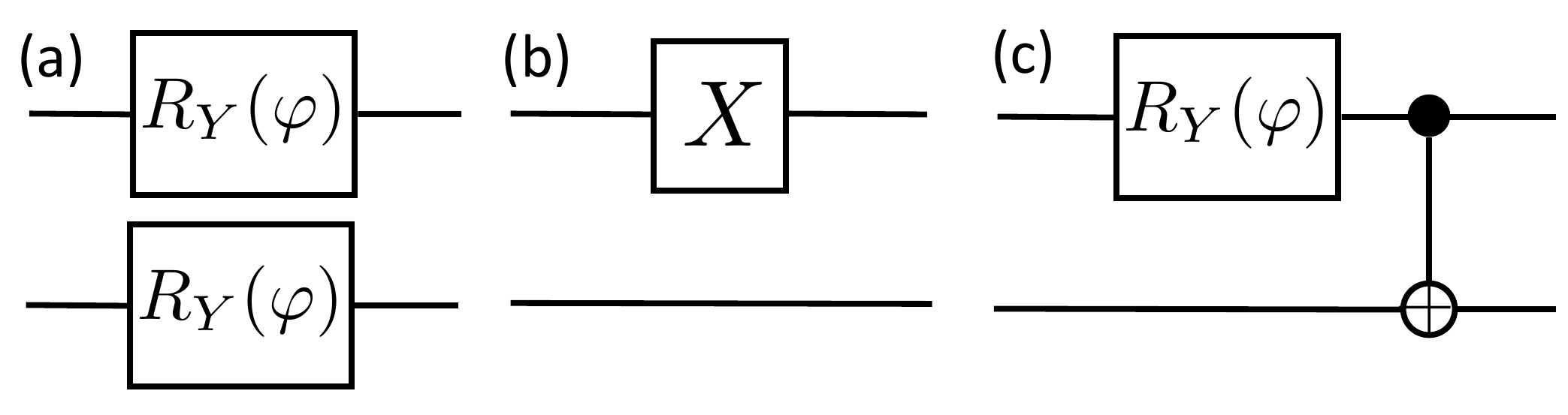}.
    \caption{Illustration of the 3 circuits used in the text to prepare a pair of time-reversed states.}
    \label{fig:3circuits}
\end{figure}

We first need to specify a convenient initial state $| \psi\rangle$. Guided by the BCS/HFB approach \cite{Bla86}, a Gaussian state 
breaking the $U(1)$ symmetry is considered: 
\begin{eqnarray}
| \psi \rangle &=& \prod_{n>0} e^{i \varphi \left( X_n Y_{n+1} + Y_n X_{n+1} \right)/2 } | -  \rangle, \label{eq:bog1}
\end{eqnarray}
where $| -\rangle =  | 0, \cdots, 0 \rangle_{n_q}$.  A general quantum circuit to obtain this state is given in \cite{Jia18} (see also \cite{Ver09}). A simpler circuit is used here noting that:
\begin{eqnarray}
 | \psi \rangle = \prod_{n}  \left[ \cos \left( \frac{\varphi}{2}\right) I_{n} \otimes I_{n+1} + \sin\left( \frac{\varphi}{2}\right) Q^+_n Q^+_{n+1} \right] | - \rangle, \label{eq:bog2}
\end{eqnarray} 
where the product is made on even $n$ only. For a given pair, the state $[\cos \left(\varphi /2 \right) | 00 \rangle +\sin\left(\varphi / 2\right)| 11 \rangle]$ is produced. This state, interpreted as a generalized Bell state, is obtained by applying the simple circuit 
shown in Fig.  \ref{fig:3circuits}-(c) to all $(n,n+1)$ pairs.   

\begin{figure}[htbp]  
\includegraphics[width=0.8\linewidth]{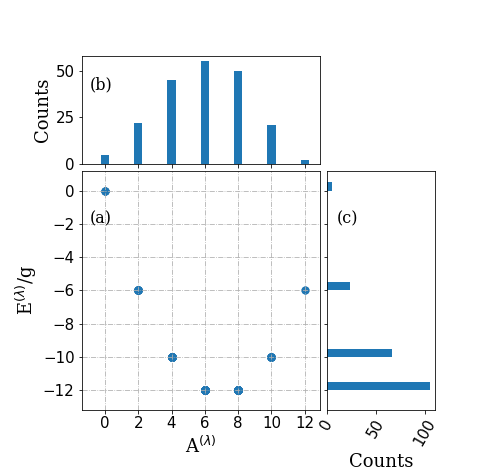} 
    \caption{Panel (a): Correlation between the energies $E^{(\lambda)}$ and particle number $A^{(\lambda)}$ obtained with the IBM Qiskit emulator \cite{Abr19} for $200$
    events using the DSA method for $n_q=12$ Qubits ($6$ pairs) with 3 register qubits and $\varphi=\pi/2$. The distribution of counts 
    for the particle number and energies are shown respectively in panels (b) and (c).}
    \label{fig:evenn06block0}
\end{figure}

The DSA approach is applied to the state (\ref{eq:bog2}) using $U_N$ as a filter. 
For each measurement, labeled by $(\lambda)$, a specific string of $0$ or $1$ is measured for the 
register qubits. The measured binary string equals $A^{(\lambda)}/2^{n_r}$ where $A^{(\lambda)}$ is the particle number of the event. 
After the measurement, a set of states $| \psi^{(\lambda)}_f \rangle$  is obtained in the working qubit basis, each of them having exactly the particle number $A^{(\lambda)}$. A hybrid calculation is then performed
by computing on a classical computer the energy $E^{(\lambda)} = \langle \psi^{(\lambda)}_f | H_P| \psi^{(\lambda)}_f\rangle $. The statistical ensemble of energies obtained in a single run is 
displayed in Fig. \ref{fig:evenn06block0}.  The ground-state (GS) energies of all even particle numbers as given by Eq. (\ref{eq:senior}) with $\nu=0$ are recovered in this run illustrating the advantage of quantum parallelism. 
For the degenerate case, unless the specific situation $\varphi = k \pi $ is considered, 
all eigenvalues are obtained from a single-value of $\varphi$ and only the 
probability distributions displayed in panels (b) and (c) of Fig. \ref{fig:evenn06block0} depend on $\varphi$. This method can be generalized to 
treat a more complex pairing Hamiltonian by allowing  $Y$-rotation with different angles $\varphi_n$ for different pairs. The set of $\{ \varphi_n\}$
can then be used as variational parameters to construct highly entangled states that can be used for instance in a VQE algorithm.   

\begin{figure}[htbp]  
\includegraphics[width=0.8\linewidth]{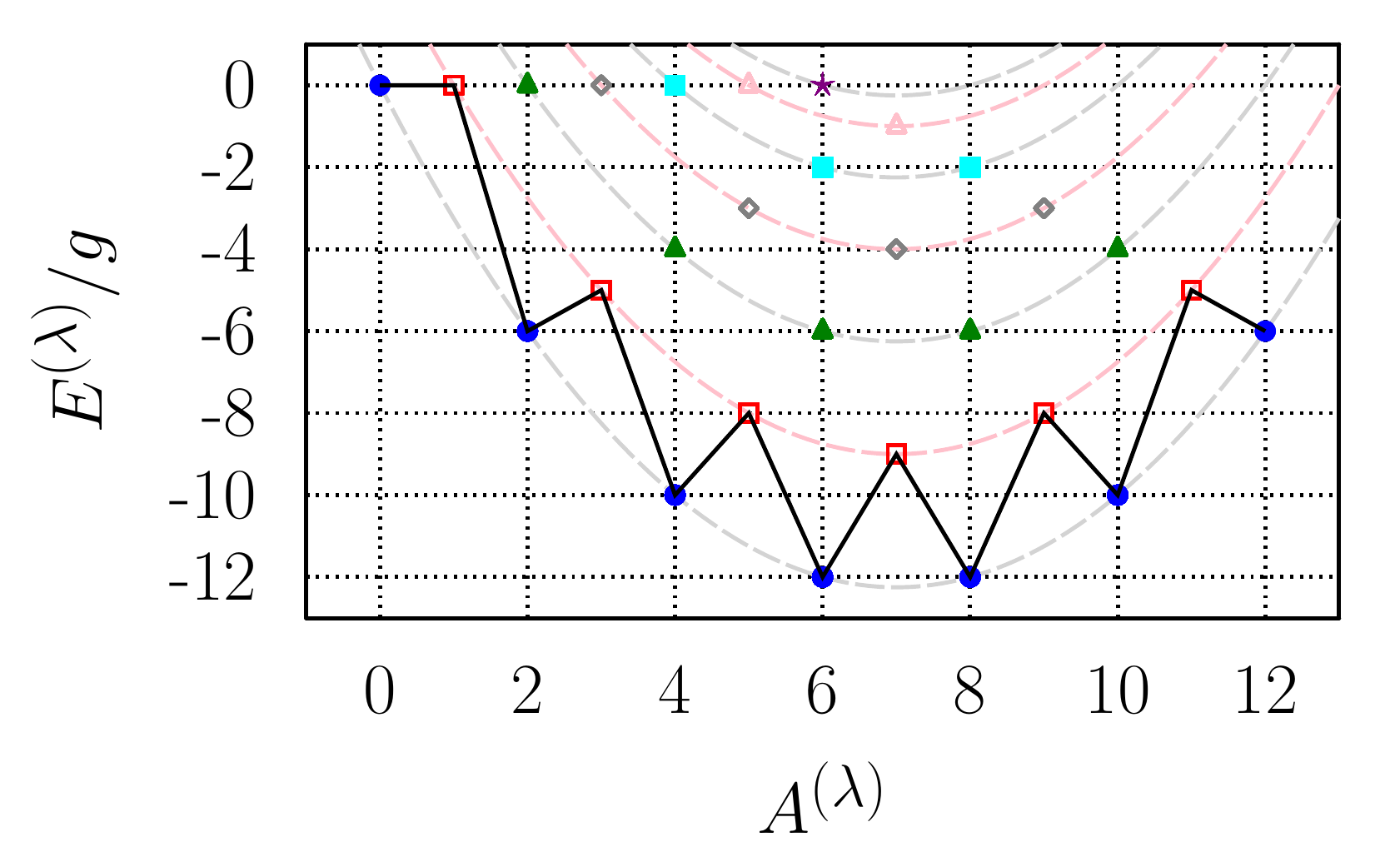} 
    \caption{Energies obtained for $n_q=12$ by replacing  the circuits (c) displayed in Fig. \ref{fig:3circuits} by the circuit (b) for an increasing number of pairs. The number of replacements then corresponds to the seniority value. 
Results have been obtained using the IBM Qiskit simulator \cite{Abr19} with $200$ events and $n_q=12$ working qubits.
 For even particle numbers, $\nu=0$ (blue filled circles), $2$ (green filled triangles), $4$ (cyan filled squares), $6$ (purple star) are shown. For odd particle numbers, $\nu=1$ (open red squares), $3$ (open gray diamond), $5$ (open pink triangle) are shown. The dashed lines correspond to the analytical result (\ref{eq:senior}) with different seniority values. Even (resp. odd) seniorities are shown with gray (resp. pink) dashed lines.  The black line connects the results obtained by replacing for one pair the circuit (c) by the circuit (a). 
    }
    \label{fig:senority}
\end{figure}
I finally show in Fig. \ref{fig:senority} that the blocking technique sometimes used in superfluid systems \cite{Rin80} can easily be transposed to qubit systems to access excited states in odd or even systems. 
One or several pairs can be broken by replacing the circuits (c) displayed in Fig. \ref{fig:3circuits} by the circuit (b). The correlations between the $A^{(\lambda)}$ and $E^{(\lambda)}$
obtained by breaking from $1$ to $n_q/2$ pairs is shown in Fig. \ref{fig:senority}. Imposing one broken pair gives the GS energy of odd systems while breaking two pairs gives the 
first excited state in even system. Breaking more and more pairs finally gives the full odd-even spectra.   
There exists a large flexibility to be explored to access selected parts of the spectra with various particle numbers. I show in Fig. \ref{fig:senority} that the GS energy of both odd and even systems can be simultaneously obtained by replacing simply for one pair the circuit (c) by the circuit (a) of Fig. \ref{fig:3circuits}.

In summary, an approach to obtain strongly entangled states that might be useful to describe interacting systems on a quantum computer is presented here. 
Starting from an operator having a known discrete spectra, the QPE approach is used to obtain an ensemble of 
entangled states. When the operator is related to a symmetry of 
the Hamiltonian, the protocol proposed here can be interpreted as a quantum computer equivalent to the 
SB-SR 
approach that is a powerful tool to describe static and dynamical properties of interacting mesoscopic systems \cite{Bla86,Rin81,Hen14,Qiu19,Gam12,Reg19,Kha20}. 
The DSA method can be applied even if the linear condition $\lambda_k = a m_k$ is relaxed by introducing the 
operator  $U_S = \displaystyle  \exp\left\{ 2\pi i (S - \lambda_0)/\Delta \right\}$ with $\Delta =\lambda_M - \lambda_0 + \varepsilon$. $\varepsilon$
is a small number insuring that the eigenvalues of $U_S$, given by $e^{i2\pi \theta_k}$, verify $\theta_k = (\lambda_k - \lambda_0)/\Delta < 1$. 
Since in general the $\{ \theta_k \}$ will not be written exactly as a truncated binary fraction, the application of the DSA approach is anticipated to require 
more register qubits in order to discriminate all channels.  One can introduce the quantity $\theta_d = d/ \Delta$ and its binary fraction $\theta_d = 0.x_1x_2 \cdots$, where $d=\min_{0 \le k < M} (\lambda_{k+1} - \lambda_k)$. Assuming that $x_m$ is the first non-zero value in the binary fraction expansion, a minimal  condition is then $n_r > m$.  Multiple projections of 
commuting operators can also be made at the price of increasing the number of register qubits. 
The restoration of broken symmetries is a pillar in the treatment of complex interacting systems beyond the perturbative regime. 
The state-of-the-art is to use it in the Variation-After-Projection (VAP) version \cite{Rin81,Ben03,Rob19} or used to propose novel  
many-body techniques \cite{Hen14,Rip17,Rip18}. In the VAP and in these new techniques, the projection becomes intractable especially when several symmetries are simultaneously  restored as it may happen for example in nuclei. The possibility to perform multiple projections on a quantum computers open perspectives in this context. 


\section*{Acknowledgments}
This project has received financial support from the CNRS through the 80Prime  program.
I thank M. Grasso, J. Carbonell, G. Hupin, F. Farget and S. Incerti for their continuous support 
in the project. I acknowledge the use of IBM Q cloud as well as use of the Qiskit software package \cite{Abr19} for performing the quantum simulations.

\end{document}